\documentclass[letterpaper, 10 pt, conference]{IEEEtran}
\newcounter{mytempeqncnt}
\newtheorem{thm}{Theorem}
\newtheorem{lem}{Lemma}

\newtheorem{crlry}{Corollary}
\newtheorem{defntn}{Definition}

\newcommand{\qed}{\nobreak \ifvmode \relax \else
      \ifdim\lastskip<1.5em \hskip-\lastskip
      \hskip1.5em plus0em minus0.5em \fi \nobreak
      \vrule height0.75em width0.5em depth0.25em\fi}

\ifCLASSINFOpdf
  \usepackage[pdftex]{graphicx}
  \graphicspath{{./}{Figs/}}
  \DeclareGraphicsExtensions{.pdf,.jpeg,.png}
\else
\fi
\usepackage{amssymb}
\usepackage{amsmath}
\usepackage{mathtools}
\usepackage{bigdelim}
\usepackage{tikz}
\usepackage{bbm}
\usepackage{mathtools}
\usepackage{centernot}
\usepackage{enumerate}
\usepackage{subcaption}
\usepackage{pgfplots}
\usepackage[caption=false,font=footnotesize]{subfig}
\usepackage{undertilde}
\usepackage{verbatim}
\usetikzlibrary{arrows,calc}
\usepackage{relsize}

\IEEEoverridecommandlockouts

\begin{document}

\title{Evolution of Social Networks: A Microfounded Model}
\author{\IEEEauthorblockN{Ahmed M. Alaa, Kartik Ahuja, and Mihaela van der Schaar}
\IEEEauthorblockA{Department of Electrical Engineering, UCLA\\
Emails: \{ahmedmalaa@ucla.edu, ahujak@ucla.edu,mihaela@ee.ucla.edu\}}}

\maketitle
\begin{abstract}
Many societies are organized in networks. Real-world social networks such as friendship networks, online social networks, scientific collaboration and citation networks, are formed by people who meet and interact over time. The way people meet is highly influenced by the evolving network structure, and their decisions to connect depend mainly on their intrinsic characteristics. In this paper, we present a first mathematical model to capture the microfoundations of social networks evolution, where people modeled as boundedly rational agents of different types join the network, meet other agents stochastically over time, and consequently decide to form a set of {\it social ties}. Based on the meeting process, which is governed by the level of {\it structural opportunism} and the a priori {\it type distribution}, as well as the incentives of the agents to form links, which are governed by {\it homophily} and {\it social gregariousness}, agents make a sequence of link formation decisions that lead to an endogenously evolving network. A basic premise of our model is that in real-world networks, agents do not form links in a one-shot fashion via a {\it preferential attachment} probabilistic rule, but rather form links by reasoning about the social benefits that agents they meet over time can bestow. We analytically study the evolution of the endogenously formed networks in terms of friendship and popularity acquisition given the following exogenous parameters: structural opportunism, type distribution, homophily, and social gregariousness. We show that the time needed for an agent to find ``friends" is highly influenced by the exogenous network parameters: agents who are more gregarious, more homophilic, less opportunistic, or belong to a type ``minority"  spend, on average, a longer time searching for friends. Moreover, we show that preferential attachment and thus, an agent's popularity acquisition, is a direct consequence of an endogenously emerging {\it preferential meeting} process, in which agents who search for friendships meet more popular agents with higher probability. We also show that the meeting process can be {\it doubly preferential}, in which agents of a certain type meet more popular similar-type agents with higher probability. Such meeting process creates asymmetries in the levels of popularity attained by different types of agents.
\end{abstract}
\IEEEpeerreviewmaketitle
\section{Introduction}

We are living in the era of networks. With the widespread usage of online social networking (OSN) platforms, such as Facebook and Twitter \cite{ref01}; academic networking websites such as ResearchGate \cite{ref02}; and professional online networks such as Linkedin \cite{ref03}, people are getting more and more connected. With people interacting and getting connected through these platforms, networks emerge endogenously as a result of the actions of people who meet others over time, and take link formation decisions, i.e. ``follow" a user on Twitter, ``add" a friend on Facebook, ``cite" a paper that is indexed by Google Scholar, or ``collaborate" with a researcher. Examples of emerging networks include: friendship networks \cite{ref10} \cite{ref19} \cite{ref39}, scientific collaboration and citation networks \cite{ref24}, and professional networks \cite{ref03}. Understanding how networks form and evolve is essential for drawing insights into networked social interactions, carrying out predictions, and designing policies that can guide network formation. While extensive research has been recently devoted to the study of social networks, no systematic model exists that can explain how networks form and evolve over time based on the individuals decisions and preferences.

\begin{figure}[t!]
    \centering
    \includegraphics[width=3.5 in]{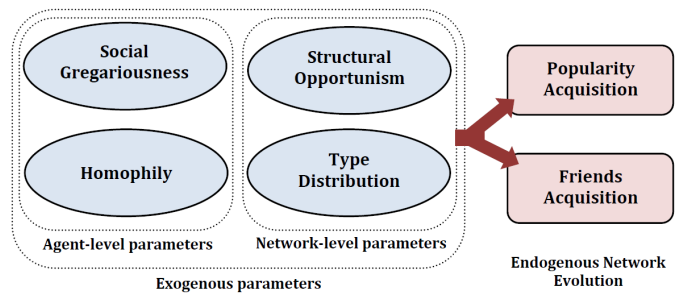}     
		\captionsetup{font= small}
    \caption{Framework for the analysis of social networks evolution.}
\end{figure}

In this paper, we present a comprehensive micro-foundational model and analysis for dynamic social network formation. In our model, networks are formed over time by the actions of boundedly rational agents that arrive at the network stochastically, and meet other agents via a random process that is itself highly influenced by the dynamic network structure and the characteristics of the agents themselves. Thus, networks evolve over time as a stochastic process driven by the individual agents, where the formation of {\it social ties} among agents are in part endogenously determined, as a function of the current network structure itself, and in part exogenously, as a function of the individual characteristics of the agents. Agents have bounded rationality, i.e. they only have information about other agents they meet over time, and they are not able to observe the global network structure or reason about links formed by others. We focus on the impact of various exogenous parameters that describe both the characteristics of individuals forming the network, and the nature of the network itself, on the endogenously evolving network structure. While many network metrics such as diameter, clustering coefficient \cite{ref5}, community structures \cite{ref24}, and degree distribution \cite{ref7} can be computed using our model, we focus on two basic aspects of network evolution that describe the agent-level experience in the network: {\it friendship acquisition} (the process of forming links) and {\it popularity acquisition} (the process of gaining links), and we show how these experiences depend on the parameters considered. Before presenting our model and results, we provide the following definitions for the exogenous parameters under study. Fig. 1 depicts and categorizes these parameters. \\
{\bf 1- Type Distribution:} Agents are heterogenous in the sense that they possess type attributes that correspond to their preferences, race, ethnicity, etc. The experiences of different types of agents in the network are generally not symmetric. The type distribution characterizes the fraction of agents of each type in the network. We say that an agent belongs to a {\it type minority} to qualitatively describe a scenario where the fraction of agents of the corresponding type in the population is small, and we say that an agent belongs to a {\it type majority} otherwise. \\
{\bf 2- Homophily:} A pervasive feature of social networks that corresponds to the tendency of the agents to connect to other similar-type agents \cite{ref381} \cite{ref93} \cite{ref90}. The extent to which a certain type of agents is {\it homophilic} is captured by an {\it exogenous homophily index}, which we formally define in section II. The homophily index can be thought of as the amount of ``intolerance" that a certain type of agents have towards making contacts with other types.\\ 
{\bf 3- Social Gregariousness:} Some types of agents can be more {\it social} than others, and thus are willing to form more links. Social gregariousness is simply measured by the minimum number of links an agent is willing to make.\\ 
{\bf 4- Structural Opportunism:} Agents in the network are said to be {\it opportunistic} if they exploit their contacts to find new contacts; thus, agents are more likely to link with the friends of their friends if they are opportunistic. Structural opportunism can also be interpreted as a {\it social norm} that agents are expected to follow. For instance, users in Twitter are expected to {\it retweet} the tweets posted by users they follow, which leads to the followers of followers of a certain user to follow him. Structural opportunism can also be a social norm in friendship networks, where people introduce their friends to each other or people enjoy/trust the friends of their friends more than strangers.\\

\subsection{Preview of the results} 
The central finding of this paper is that an agent's experience in the network is affected by the agents it meets over time. This meeting process is in turn affected by the exogenous network parameters. We classify our results as follows:\\ 
{\bf 1- Friendship acquisition characterization:} Agents joining the network will form a finite number of links over time. We say that agent A is a {\it friend} of agent B if agent B forms a directed link with agent A. In section III-A, we study the impact of the exogenous parameters on the time needed for an agent to find its {\it friends}. We show that agents who are more gregarious, more homophilic, less opportunistic, or belong to a type ``minority" are more likely to spend more time searching for friendships. \\
{\bf 2- Popularity acquisition characterization:} After an agent joins the network, other agents form links with it. The number of such links reflects the level of popularity of that agent, e.g. number of followers in Twitter, and number of citations in a citation network. We say that agent A is a {\it follower} of agent B if agent A forms a directed link with agent B. In section III-B, we study the impact of the exogenous parameters on the popularity acquisition time and popularity growth rates in a large network. We show that popularity evolution depends on the meeting process, which can in general be {\it doubly preferential}, i.e. the meeting process is both type and popularity biased. We prove that depending on the exogenous parameters, preferential attachment, which corresponds to the cumulative advantage in popularity acquisition, emerges as a result of the preferential meeting process. Moreover, we show that in homophilic societies, an emerging doubly preferential meeting process allows more gregarious agent types to get more popular, while in non-homophilic societies, an agent's age in the network determines its popularity over time.

\subsection{Related works}
Previous works on network formation can be divided into three categories: networks formed based on {\it random events} \cite{ref9}, \cite{ref90}-\cite{ref13}, networks formed based on {\it strategic decisions} \cite{ref91}-\cite{ref17}, and empirical models distilled by mining networks' data \cite{ref10}-\cite{ref24}, \cite{ref36}. A fairly large body of literature has been devoted to developing mathematical models for network formation, yet much fewer works attempt to interpret and understand how networks evolve over time and how can individual agents affect the characteristics of such networks. Probabilistic models based on random events are generative models that are concerned with constructing networks that mimic real-world social networks. In \cite{ref90}-\cite{ref18}, agents get connected in a pure probabilistic manner in order to realize some degree distribution \cite{ref3}, or according to a {\it preferential attachment} rule \cite{ref4} \cite{ref5}. While such models can capture the basic structural properties of social networks, they fail to explain why and how such properties emerge over time. In contrast, strategic network formation models such as those in \cite{ref91}-\cite{ref17}, and our previous works in \cite{ref25} \cite{ref16}, can offer an explanation for why certain network topologies emerge as an equilibrium of a network formation game. However, these results are limited to studying network {\it stability} and {\it efficiency}, and provide only very limited insight into the dynamics and evolution of networks. Finally, mining empirical data can help building algorithms for detecting communities \cite{ref20}-\cite{ref361}, predicting agents' popularity \cite{ref36}, or identifying agents in a network \cite{ref31}, but cannot help understanding how networks form and evolve. 

\section{Model}
\subsection{Network model}
We consider a discrete-time model for a growing social network where one {\it agent} is born each time step and is indexed by its birth date $i \in \{1,2,. \, . \, ., t,. \, . \, .\}$. At date $t \in \mathbb{N}$, a snapshot of the network is modeled by a {\it step graph} $\mathcal{G}^{t}$ given by $\mathcal{G}^{t} = (\mathcal{V}^{t}, \mathcal{E}^{t})$,  where $\mathcal{V}^{t}$ is the set of nodes, $\mathcal{E}^{t} = \{e^{t}_{1}, e^{t}_{2},.\,.\,.,e^{t}_{|\mathcal{E}^{t}|}\}$ is the set of edges between different nodes, with each edge $e^{t}_{k}$ being an ordered pair of nodes $e^{t}_{k} = (i,j)$ $\left(i \neq j, \, \mbox{and} \, i,j \in \mathcal{V}^{t}\right)$, and $|\mathcal{E}^{t}|$ is the number of distinct edges in the graph. Thus, $\mathcal{G}^{t}$ is a directed graph. Nodes correspond to agents (social actors) and edges correspond to directed links (social ties) between the agents. The adjacency matrix of $\mathcal{G}^{t}$ is denoted by ${\bf A}^{t}_{\mathcal{G}} = [A^{t}(i,j)], A^{t}(i,j) \in \{0,1\}, A^{t}(i,i) = 0, \forall i,j \in \mathcal{V}^{t}$. An entry of the adjacency matrix $A^{t}(i,j) = 1$ if $(i,j) \in \mathcal{E}^{t}_{k}$, and $A^{t}(i,j) = 0$ otherwise. If $A^{t}(i,j)=1$, then agent $i$ initiates a link with agent $j$, and we say that $j$ is a ``friend" of $i$, and $i$ is a ``follower" of $j$. The directed nature of a link indicates the agent initiating the link, and only this agent obtains the {\it social benefit} of linking and pays the link cost. The {\it indegree} of agent $i$ is the number of links that are initiated towards $i$, denoted by $\mbox{deg}^{-}_{i}(t)$, while the {\it outdegree}, denoted by $\mbox{deg}^{+}_{i}(t)$, is the number of links initiated by agent $i$. 

Each agent $i \in \mathcal{V}^{t}$ in the network possesses a type attribute $\theta_{i}$, which belongs to a finite set of types $\theta_{i} \in \Theta, \Theta = \{1,2,3,.\,.\,.,|\Theta|\}$, where $|\Theta|$ is the number of types. The set of type-$k$ agents at time $t$ is denoted by $\mathcal{V}_{k}^{t}$, where $\mathcal{V}^{t} = \bigcup_{k =1}^{|\Theta|} \mathcal{V}_{k}^{t}$. Agents are identified by both their birth dates and types. The network starts initially with a seed graph $\mathcal{G}^{0}$, which we assume to be an empty graph with no agents at $t = 0$, and agents arrive one at a time at each date $t$. The agents arrival in the network is modelled as a stationary stochastic process $\lambda(t) = \{\theta_{t}\}_{t\in\mathbb{N}}$, with a sample space $\Lambda = \Theta^{\mathbb{N}}$, i.e. $\Lambda = \left\{\left(\theta_{1}, \theta_{2}, .\,.\,.\right): \theta_{t} \in \Theta, \, \forall t \in \mathbb{N}\right\}$. We assume that the types of agents are independent and identically distributed, and that the agents' {\it type distribution} is $\mathbb{P}(\theta_{i} = k) = p_{k}$, where $\sum_{k \in \Theta} p_{k} = 1$. Thus, $\lambda(t)$ is a {\it Bernoulli scheme}. Using {\it Borel's law of large numbers}, we know that 
\[\mathbb{P}\left(\lim_{t \rightarrow \infty} \frac{1}{t}  \left|\mathcal{V}_{k}^{t}\right| = p_{k}\right) = 1.\]
In other words, for a sufficiently large network size (and age $t$), the actual fraction of agents of each type in the network converges almost surely to the prior type distribution of the Bernoulli scheme. Thus, at date $t$, and for a large enough network, the expected number of type $k$ agents in the network is $p_{k}t$, and the total number of agents is $t$, i.e. $|\mathcal{V}^{t}| = t,$ $\mathbb{E}\left\{|\mathcal{V}_{k}^{t}|\right\} = p_{k}t$, and $\lim_{t \rightarrow \infty} \frac{|\mathcal{V}_{k}^{t}|}{|\mathcal{V}^{t}|} = p_{k}$. 

\subsection{The meeting process}
Let $M_{i}(t) = \{m_{i}(t)\}_{t = i}^{i+T_{i}-1}$ be the {\it meeting process} of agent $i$, which corresponds to the sequence of birth dates of the agents that agent $i$ meets over time (note that birth dates are used to identify agents), and $T_{i}$ is the stopping time of $M_{i}(t)$, which we define in section III-A as the {\it link formation time}. The sample space of the meeting process is given by $\mathcal{M} = \left\{\left(m_{i}(i), m_{i}(i+1), .\,.\,., m_{i}(i+T_{i}-1)\right): m_{i}(t) \leq t, m_{i}(t) \neq i\right\}$. Unlike the arrival process, which is stationary and exogenous, the meeting process depends on the history of actions of all agents in the network, i.e. the probability that agent $i$ meets agent $j$ at time $t$ depends on their relative positions in the network at time $t$, which in turn depend on the sequence of meetings for both agents up to time $t-1$. Moreover, the probability that a certain sample path of the meeting process occurs depends on all the exogenous parameters shown in Fig. 1. We denominate the set of agents to whom agent $i$ forms links as agent $i$'s {\it friends}, and the set of agents that form links with agent $i$ by agent $i$'s {\it followers}. Denote the set of type-$k$ friends of agent $i \in \mathcal{V}^{t}$ by $\mathcal{N}^{+,k}_{i,t}$, and the set of all friends of $i$ as $\mathcal{N}_{i,t}^{+} = \bigcup_{k=1}^{|\Theta|} \mathcal{N}^{+,k}_{i,t}$, where $|\mathcal{N}^{+}_{i,t}| = \mbox{deg}_{i}^{+}(t)$. Similarly, we denote the followers of agent $i$ by $\mathcal{N}_{i,t}^{-}$, where $|\mathcal{N}^{-}_{i,t}| = \mbox{deg}_{i}^{-}(t)$. Define the set $\mathcal{K}_{i,t} = \bigcup_{j \in \mathcal{N}^{+}_{i,t-1}} \mathcal{N}^{+}_{j,t-1}/\left\{i\right\}$ as the set of {\it friends of friends} of agent $i$ at time $t$, and the set $\bar{\mathcal{K}}_{i,t} = \mathcal{V}^{t}/\left\{\mathcal{K}_{i,t} \bigcup \mathcal{N}^{+}_{i,t} \bigcup i \right\}$ as the set of {\it strangers} to agent $i$ at time $t$. We capture the degree of structural opportunism of agents in the society by a parameter\footnote{We assume that all types of agents have the same $\gamma$. The analysis can be easily extended to the case when each type has a different $\gamma$.} $\gamma \in [0,1]$, where $\gamma = 0$ corresponds to fully opportunistic agents, and $\gamma = 1$ corresponds to non-opportunistic agents. That is, $\gamma$ is a measures of how often an agent $i$ finds new friends without exploiting its current connections as we show in the following meeting process. For $t \geq i$, agent $i$ meets {\it one} agent selected uniformly at random from the set of friends of friends with probability $1-\gamma$ if $\mathcal{K}_{i,t} \neq \phi$, while if $\bar{\mathcal{K}}_{i,t} = \phi$, then agent $i$ meets one agent selected from $\mathcal{K}_{i,t}$ with probability 1, i.e.
\[\mathbb{P}\left(m_{i}(t) \in \mathcal{K}_{i,t}\left|\mathcal{K}_{i,t}\neq \phi, \bar{\mathcal{K}}_{i,t}\neq \phi\right.\right) = 1-\gamma,\] 
\[\mathbb{P}\left(m_{i}(t) \in \mathcal{K}_{i,t}\left|\mathcal{K}_{i,t}\neq \phi, \bar{\mathcal{K}}_{i,t} = \phi\right.\right) = 1.\]  
On the other hand, agent $i$ meets one agent selected uniformly at random from the set of strangers with probability $\gamma$ if $\bar{\mathcal{K}}_{i,t} \neq \phi$, while if $\mathcal{K}_{i,t} = \phi$, then agent $i$ meets one agent selected from $\bar{\mathcal{K}}_{i,t}$ with probability 1, i.e. 
\[\mathbb{P}\left(m_{i}(t) \in \bar{\mathcal{K}}_{i,t}\left|\mathcal{K}_{i,t}\neq \phi, \bar{\mathcal{K}}_{i,t}\neq \phi\right.\right) = \gamma,\] 
\[\mathbb{P}\left(m_{i}(t) \in \bar{\mathcal{K}}_{i,t}\left|\mathcal{K}_{i,t}= \phi, \bar{\mathcal{K}}_{i,t}\neq \phi\right.\right) = 1.\]  
Moreover, other agents can occasionally meet agent $i$ at each time step, i.e. $\mathbb{P}\left(m_{j}(t) = i\left|j \in \mathcal{V}^{t}/\{i\}, t \leq j+T_{j}-1\right.\right) > 0, \forall t \geq i$.  
   
\subsection{Agents' actions and utility functions}
When agent $i$ meets agent $m_{i}(t)$ at time $t$, it observes its type and decides whether or not to form a link with that agent. Agents draw social benefits by connecting to others, but those benefits are type-dependent and link formation is costly. Links are formed unilaterally and only the agent initiating the link bears a cost of $c$ and attains the linking benefit. We assume {\it local externalities}, i.e. linking benefits do not flow to indirect contacts. Let the benefit attained by agent $i$ from linking to agent $j$ be $\alpha_{\theta_{i}\theta_{j}}$, where $\alpha_{\theta_{i}\theta_{j}} \in \mathbb{R}^{+}$. The action of node $i$ at time $t$ is $a_{i}^{t} \in \{0,1\}$, where $a_{i}^{t} = 1$ indicates that node $i$ forms a link with node $m_{i}(t)$. The utility function of agent $i$ at date $t$ is given by 
\begin{equation}
u^{t}_{i}\left({\bf a}_{i}^{t}\right) =  v_{i}\left(\sum_{j \in \mathcal{N}^{+}_{i,t}} \alpha_{\theta_{i}\theta_{j}}\right) - c \sum_{k=i}^{t}a^{k}_{i}, 
\label{eq2}
\end{equation}
where ${\bf a}_{i}^{t} = \left(a_{i}^{i},a_{i}^{i+1},.\,.\,.,a_{i}^{t}\right)$, and $v(x): x \rightarrow \mathbb{R}^{+}$ is the benefit function that measures the {\it social capital} \footnote{Our definition for social capital follows that by Bourdieu in \cite{ref370}.}. We assume that $v(x)$ is concave\footnote{While we assume concavity of the utility function, our analysis applies to any saturating function, e.g. the {\it sigmoid} function.}, twice continuously differentiable, and monotonically increasing in $x$, and $v(0) = 0$. That is, the marginal benefit of forming links diminishes as the number of links increases. This corresponds to the fact that agents do not form an infinite number of links in the network, but rather form a ``satisfactory" number of links \footnote{For instance, in citation networks, the number of references cited in a paper is finite and corresponds to the number of papers the authors need to acquire knowledge, yet the number of citations on a specific paper can be arbitrary large.}. The action taken by agent $i$ depends on the marginal benefit of forming a link and the linking cost as shown in (\ref{eq3}), where agent $i$ forms a link to the agent it meets only if the marginal utility from linking is positive. Thus, agents are {\it myopic} and form links without taking future meetings into account. 
\begin{figure*}[!t]
\setcounter{mytempeqncnt}{\value{equation}} \setcounter{equation}{1}
\begin{equation}
a^{t}_{i} = \frac{1}{2} \, \mbox{sgn}\left(v\left(\sum_{j \in \mathcal{N}^{+}_{i,t-1}} \alpha_{\theta_{i}\theta_{j}} + \alpha_{\theta_{i} \theta_{m_{i}(t)}}\right) - v\left(\sum_{j \in \mathcal{N}^{+}_{i,t-1}} \alpha_{\theta_{i}\theta_{j}}\right) - c\right)+\frac{1}{2}.
\label{eq3}
\end{equation}
\setcounter{equation}{\value{mytempeqncnt}+1} \hrulefill{}\vspace*{4pt}
\end{figure*}
To capture the impact of {\it homophily}, we assume that $\alpha_{\theta_{i}\theta_{k}} > \alpha_{\theta_{i}\theta_{j}}, \forall \, \left|\theta_{k}-\theta_{i}\right| < \left|\theta_{j}-\theta_{i}\right|$, and $\alpha_{\theta_{i}\theta_{k}} = \alpha_{\theta_{i}\theta_{j}}, \forall \, \left|\theta_{k}-\theta_{i}\right| = \left|\theta_{j}-\theta_{i}\right|$. 
\subsection{The exogenous parameters}
In order to measure the exogenous homophilic tendency of a certain type of agents, we propose a novel definition of an {\it exogenous homophily index} for type-$k$ agents $h_{k}$, which is a variant of the well known {\it Coleman homophily index} \cite{ref13}:
\begin{align}
h_{k} &\triangleq 1-\frac{\sum_{m \neq k}^{|\Theta|} p_{m} \lim_{t \rightarrow \infty} r_{\theta_{i}}(m,t)}{1 - p_{k}}, \forall \theta_{i}=k
\label{eq221}
\end{align}
where $r_{\theta_{i}}(m,t)$ is the {\it maximum excess representation} of type-$m$ agents in agent $i$'s friends at time $t$, which is given by
\begin{equation}
r_{\theta_{i}}(m,t) \triangleq \sup_{\mathcal{N}^{+}_{i,t}} \frac{\left|\mathcal{N}^{+,m}_{i,t}\right|}{\mbox{deg}_{i}^{+}(t)},
\label{eq3XX}
\end{equation}
where $r_{\theta_{i}}(k,t) = 1, \forall \theta_{i} = k$, and $0 \leq h_{k} \leq 1, \forall k \in \Theta$. When type-$k$ agents are indifferent to the types of agents it connects to, i.e. type-$k$ agents are extremely non-homophilic, then we have $\lim_{t \rightarrow \infty} r_{\theta_{i}}(m,t) = 1, \forall \theta_{i} = k, m \in \Theta,$ which means that $h_{k} = 0$. On the other hand, if agents restrict their links to same-type agents only, then $\lim_{t \rightarrow \infty} r_{\theta_{i}}(m,t) = \delta\left(m,k\right), \forall \theta_{i} = k, m \in \Theta,$ where $\delta\left(x,y\right)$ is the {\it Kronecker delta} function, which means that $h_{k} = 1$. Now we compute the exogenous homophily index in closed-form. Let $L_{\theta_{i}}^{*}(k,\alpha) \in \mathbb{Z}$ be the maximum number of links with type-$k$ agents that agent $i$ can form in the time period $[T,\infty)$ given that $\sum_{j \in \mathcal{N}^{+}_{i,T-1}} \alpha_{\theta_{i}\theta_{j}} = \alpha$, i.e. $L_{\theta_{i}}^{*}(k,\alpha) \triangleq \sup \left(\lim_{t\rightarrow \infty}\left|\mathcal{N}^{+,k}_{i,t}\right|-\left|\mathcal{N}^{+,k}_{i,T-1}\right|\right)$, for $\sum_{j \in \mathcal{N}^{+}_{i,T-1}} \alpha_{\theta_{i}\theta_{j}} = \alpha$. This can be easily computed in closed-form by taking the first derivative of the inverse of the concave benefit function in (\ref{eq2}). It can be shown that
\[\inf_{M_{i}(t) \in \mathcal{M}} \lim_{t \rightarrow \infty} \mbox{deg}_{i}^{+}(t) = L_{\theta_{i}}^{*}\left(\theta_{i},0\right),\]      
and the exogenous homophily index of agent $i$ is given by\footnote{A detailed proof can be found in \cite{ref37X}.}
\[h_{\theta_{i}} = \frac{1}{1-p_{\theta_{i}}}\left(1-\sum_{k = 1}^{|\Theta|} p_{k} \frac{L_{\theta_{i}}^{*}(k,0)}{L_{\theta_{i}}^{*}(k,0)+L_{\theta_{i}}^{*}(\theta_{i}, \alpha_{\theta_{i}k}\, L_{\theta_{i}}^{*}(k,0))}\right).\]
The parameter $L_{\theta_{i}}^{*}\left(\theta_{i},0\right)$ represents the minimum number of links an agent can form; this parameter captures social gregariousness. In summary, our model captures the four exogenous parameters defined in section I as follows:
\begin{itemize} 
\item {\it Homophily}: the homophily of type-$k$ agents is captured by the exogenous homophily index $h_{k}$.  
\item {\it Social gregariousness}: the gregariousness of type-$k$ agents is captured by $L_{k}^{*}(k,0)$. 
\item {\it Structural opportunism}: the parameter $\gamma$ reflects the extent of structural opportunism.  
\item {\it Type distribution}: the fraction of type-$k$ agents in a large network is given by $p_{k}$.
\end{itemize}

In the next section, we study the friendships acquisition experience for agents in the network. {\bf Due to space limitations, the proofs for all the Lemmas, Theorems and Corollaries in this paper are provided in the online appendix in \cite{ref37X}.}   

\section{Friendships acquisition: how long does it take to find friends?}
In the next subsection, we characterize the time needed for agents to find their friends in the evolving network. Unlike previous works where link formation is a one-shot process (which is the case in \cite{ref9}, \cite{ref5}, \cite{ref18}, \cite{ref101}, \cite{ref12}, \cite{ref13}, \cite{ref91}, \cite{ref92}, \cite{ref15}, and \cite{ref17}), in our model agents form links over time, and all agents can meet other agents and take link formation actions at every time step as long as their utility functions are not yet saturated. Based on such dynamic model, a distinguishing characteristic of a network is the time span over which agents keep forming links, i.e. how long would an agent keep searching for friends upon its arrival. Based on the definition of the utility function in (\ref{eq2}) and (\ref{eq3}), we know that there exists a time after which an agent stops forming links, which follows from our assumption of the concavity of the utility function. Moreover, the minimum number of friends that an agent makes in the network reflects the agent's gregariousness. The time horizon over which the agent forms this number of links is a function of all the exogenous parameters since it clearly depends on the meeting process, i.e. the time span over which an agent forms links is random as it depends on the types of agents that the agent meets over time and the history of the link formation decisions. For an agent $i$, the time span of link formation $T_{i}$ is defined as
\begin{equation}
T_{i} \triangleq \inf \left\{t \in \mathbb{N}: a_{i}^{\tau} = 0, \forall \tau > t\right\} - i + 1.
\label{eq8}
\end{equation}
Since $T_{i}$ is random, we characterize the time spent by an agent in the link formation process in terms of the expectation of $T_{i}$. Note that $T_{i}$ can be thought of as the {\it stopping time} of the meeting process $M_{i}(t)$. This can be easily proven by showing that the event $T_{i} = T$ only depends on the history of meetings and link formation decisions up to time $T$. We define the Expected Link Formation Time (ELFT) $\overline{T}_{i}$ as 
\begin{equation}
\overline{T}_{i} = \mathbb{E}\left[T_{i}\left|i,\theta_{i}\right.\right],
\label{eq9}
\end{equation}
where the expectation is taken over the probability mass function (pmf) of $T_{i}$, which we denote by $f_{T_{i}}(T_{i})$. In the following Theorem, we compute the ELFT for extreme cases of agents' homophily.
\begin{thm}
{\it (Homophily induces uncertainity)} For an agent $i$ born in an asymptotically large network ($i \rightarrow \infty$), if $h_{k}=0, \forall k \in \Theta$, then the LFT for agent $i$ and is equal to 
\[T_{i} = L_{\theta_{i}}^{*}(\theta_{i},0)\]
almost surely, while if $h_{k}=1, \forall k \in \Theta$, then the ELFT is given by
 \[\overline{T}_{i} = \frac{1}{p_{\theta_{i}}} + \frac{L_{\theta_{i}}^{*}(\theta_{i},\alpha_{\theta_{i}\theta_{i}})}{\gamma p_{\theta_{i}}+(1-\gamma)}. \, \, \, \, \IEEEQEDhere\] 
\end{thm}

Thus, when the agents are {\it not} homophilic, there is no uncertainity in the friendships acquisition process, and both the number of links and the link formation time are deterministic. This deterministic LFT is independent on the network, and only depends on the agent's gregariousness. That is, if $h_{k}=0, \forall k \in \Theta$, then an agent's journey in the network is controlled by the agent itself and how it values friendships, and not by the network structure or the actions of others. If agents value friendships more, i.e. are more gregarious, then they will spend more time making friends, yet this time is deterministic and only depends on parameters that are determined by the agent and not the network. On the other hand, if agents are extremely homophilic, then the agent's journey in the network will entail more stochasticity, i.e. agents are less certain about the time needed to form links since they can meet different-type agents with which they do not form any links. It is clear from Theorem 1 that the ELFT of extremely homophilic agents depends on the type distribution and opportunism, in addition to gregariousness. We emphasize these dependencies in the following corollary. 
\begin{crlry}
{\it (Gregarious agents and minorities search for friends longer, opportunistic agents search shorter)} If $h_{k}=1, \forall k \in \Theta$, then for an agent $i$, the ELFT is:
\begin{itemize}
\item a monotonically increasing function agent $i$'s gregariousness $L_{\theta_{i}}^{*}(\theta_{i},0)$.
\item a monotonically decreasing function of $p_{\theta_{i}}$, and 
\item a monotonically increasing function of $\gamma$.  \, \IEEEQEDhere 
\end{itemize} 
\end{crlry}

\begin{figure}[t!]
    \centering
    \includegraphics[width=3 in]{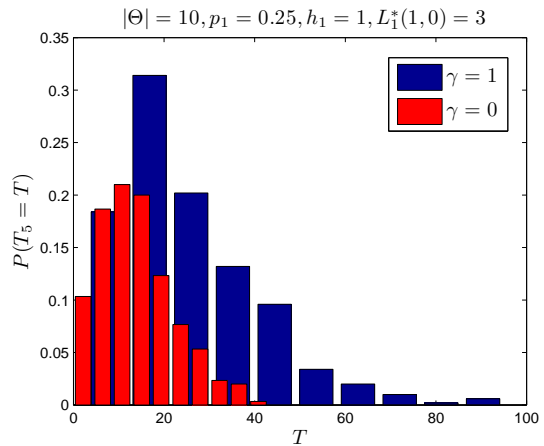}     
		\captionsetup{font= small}
    \caption{Stochastic dominance of the LFT with decreasing structural opportunism.}
\end{figure}
Corollary 1 says that the ELFT is monotonic in gregariousness, which is intuitive since the more friends an agent is willing to make, the longer it takes to find those friends. Moreover, agents belonging to minorities are expected to spend more time in the link formation process. This is again intuitive since when the fraction of similar-type agents in the population is small, each agent would need to meet a longer sequence of agents in order to find  similar-type friends. Finally, the ELFT decreases as structural opportunism increases. This is because once the agent becomes attached to a network component of similar-type agents, it is then better to be opportunistic and keep meeting the friends of friends who are guaranteed to be similar-type agents, rather than meeting strangers with uncertain types. 

Note that the results in Theorem 1 are concerned with extreme cases of homophily, i.e. $h_{k}=0$ and $h_{k}=1$. Computing the exact ELFT for an arbitrary exogenous homophily index is not mathematically tractable due to the combinatorial nature of the agents' interactions. However, in the following Theorem, we generalize Theorem 1 and Corollary 1 using {\it stochastic ordering}\footnote{The definition for stochastic dominance can be found in section 4.5.5 in \cite{ref9}.}. 
\begin{thm}
{\it (Stochastic ordering of the LFT statistics with respect to exogenous parameters)} In an asymptotically large network, for any agent $i$ with an exogenous parameters tuple $(p_{\theta_{i}},h_{\theta_{i}},\gamma, L_{\theta_{i}}^{*}(\theta_{i},0))$, and a corresponding pmf of the LFT $f_{T_{i}}\left(T_{i}\right)$, we have:
\begin{itemize}
\item If $\tilde{p}_{\theta_{i}} > p_{\theta_{i}}$ and all other exogenous parameters are the same, then $f_{T_{i}}\left(T_{i}\right)$ {\it first-order stochastically dominates} $\tilde{f}_{T_{i}}\left(T_{i}\right)$.     
\item If $\tilde{h}_{\theta_{i}} > h_{\theta_{i}}$ and all other parameters are the same, then $\tilde{f}_{T_{i}}\left(T_{i}\right)$ first-order stochastically dominates $f_{T_{i}}\left(T_{i}\right)$.     
\item If $\tilde{\gamma} > \gamma$ and all other parameters are the same, then $\tilde{f}_{T_{i}}\left(T_{i}\right)$ first-order stochastically dominates $f_{T_{i}}\left(T_{i}\right)$. \, \, \IEEEQEDhere    
\end{itemize}  
\end{thm}
Theorem 2 generalizes Theorem 1 and Corollary 1 in the sense of stochastic ordering; the monotonicity of the ELFT as a function of the exogenous parameters, as well as the expectation of any other increasing function over the LFT pmf, follows the same behavior in Corollary 1. Note that while the ELFT can be shown to be increasing with an agent's gregariousness, the social gregariousness is actually coupled with the homophily index, thus one cannot change the gregariousness while fixing a homophily index. Fig. 2 shows that the pmf of the LFT for $\gamma = 1$ stochastically dominates that for $\gamma = 0$.
      
\section{Popularity acquisition: how popular can an agent become?}

Each agent forms a finite number of links that satisfies its social gregariousness, but the number of links that an agent receives (which quantifies the agent's popularity) can be arbitrarily large depending on the agent's type, age, and position in the network. In this subsection, we characterize the evolution of the agents' popularity over time. While the ELFT measures the time span over which the agent forms links, we introduce another measure for the time span over which an agent attains a certain level of popularity, which we term the Expected Popularity Acquisition Time (EPAT). Let $T^{p}_{i}(d)$ be the {\it popularity acquisition time}, i.e. the time period over which an agent's indegree becomes $d$, i.e.
\begin{equation}
T^{p}_{i}(d) \triangleq \inf \left\{t \in \mathbb{N}: \mbox{deg}_{i}^{-}(\tau) \geq d, \forall \tau > t\right\} - i + 1.
\label{eq112}
\end{equation}        
Note that $T^{p}_{i}(d)$ can be thought of as the difference between the {\it hitting time} of the process $\left\{\mbox{deg}_{i}^{-}(t)\right\}_{t=i}^{\infty}$ and the birth date of agent $i$. Since such period is random, we define the {\it expected popularity acquisition time} (EPAT) $\overline{T}^{p}_{i}(d)$ as
\begin{equation}
\overline{T}^{p}_{i}(d) = \mathbb{E}\left[T^{p}_{i}(d) \left|i, \theta_{i}\right.\right].
\label{eq113}
\end{equation}  
The popularity of agent $i$ at time $t$ is simply given by $\mbox{deg}_{i}^{-}(t)$. We say that the {\it popularity growth rate} is $O\left(g(t)\right)$ if $\lim_{t \rightarrow \infty}\left|\frac{\mathbb{E}\left\{\mbox{deg}_{i}^{-}(t)\right\}}{g(t)}\right| = b,$ where $b$ is a positive constant. Finally, we define the {\it popularity crossover time} $T^{p}_{c}(i,j,\theta_{i}, \theta_{j}, \gamma, \gamma^{'}),$ as the time at which the expected popularity of an agent $j$ of type $\theta_{j}$ in a network realization with structural opportunism $\gamma^{'}$ exceeds the popularity of an agent $i$ of type $\theta_{i}$ in a network realization with structural opportunism $\gamma$, i.e. 
\[T^{p}_{c}(i,j,\theta_{i}, \theta_{j}, \gamma, \gamma^{'}) \triangleq \]
\begin{equation}
\inf \left\{t \in \mathbb{N}: \mathbb{E}\left\{\mbox{deg}_{i}^{-}(\tau)\left|\gamma\right.\right\} < \mathbb{E}\left\{\mbox{deg}_{j}^{-}(\tau)\left|\gamma^{'}\right.\right\}, \forall \tau > t\right\}.
\label{eq113XX}
\end{equation}
When $\gamma = \gamma^{'}$, we simply denote the popularity crossover time by $T^{p}_{c}(i,j,\theta_{i}, \theta_{j}).$ Next, in the following, we define a new notion of {\it doubly preferential meeting processes}, which plays a central role in the popularity evolution process. 
\begin{defntn} 
({\it Doubly preferential meeting process}) We say that the meeting process $M_{i}(t)$ is {\it doubly preferential} if  
\[\mathbb{P}\left(m_{i}(t) = j \left|i \leq t \leq i+T_{i}-1, \mbox{deg}_{j}^{-}(i)\right.\right) = y_{ij}\left(\mbox{deg}_{j}^{-}(i)\right),\]
where $y_{ij}(x)$ is a linear function of $x$, and $y_{ij}(x) > y_{ik}(x), \forall x$ if and only if $p_{\theta_{j}} r_{\theta_{i}}(\theta_{j},\infty) > p_{\theta_{k}} r_{\theta_{i}}(\theta_{k},\infty)$. \IEEEQEDhere 
\end{defntn}
Thus, a doubly preferential meeting process is a process that leads agent $i$ to meet an agent $j$ with a probability proportional to both the current popularity level of agent $j$, and the maximum excess representation of type-$\theta_{j}$ agents in the friends of type-$\theta_{i}$ agents. The meeting process is {\it doubly preferential} because it gives an advantage to the more popular agents, and in addition gives an advantage to similar-type agents. For an extremely homophilic agent $i$ ($h_{\theta_{i}} = 1$), we have $y_{ij}(x) = 0, \forall x, \theta_{i} \neq \theta_{j}$. It is worth mentioning that since doubly preferential meetings allows similar-type agents to meet with higher probability, it then captures what Mayhew calls ``structuralist" homophily effects in \cite{ref93}, and what Kossinets and Watts refer to as ``induced homophily" in \cite{ref90}, together with the linear preferential attachment growth model. In the following Lemma, we show that structural opportunism and homophily promotes doubly preferential meeting processes.
\begin{lem}
{\it (Structural opportunism and homophily promote doubly preferential meetings)} For any network with $\gamma \in [0,1)$ and $h_{k} \in (0,1], \forall k \in \Theta$, the meeting process of any agent $i$ is doubly preferential.  \IEEEQEDhere  
\end{lem}
In the following Theorem, we show that {\it preferential attachment} in the popularity evolution process emerges over time as a result of the doubly preferential meeting processes, which in turn results from structural opportunism. 
\begin{thm}
{\it (Emergence of preferential attachment due to structural opportunism)} For any network with $\gamma \in [0,1)$, preferential attachment emerges over time and we have
\[\mathbb{P}\left(A^{t}(i,j) = 1 \left| \, \mbox{deg}_{j}^{-}(i), i\leq t\leq i+T_{i}-1\right.\right)  \propto y_{ij}\left(\mbox{deg}_{j}^{-}(i)\right). \] \IEEEQEDhere
\end{thm}
This Theorem says that if agents are opportunistic, then the popularity of each agent exhibits an {\it accumulated advantage} pattern where the {\it popular gets more popular}. Unlike \cite{ref4} \cite{ref5} \cite{ref101} \cite{ref371}, in our model preferential attachment emerges endogenously over the link formation time as a result of the meeting process and the actions of the agents rather than being a probabilistic link formation rule that specifies a one-shot linking behavior. Moreover, the probability that agent $i$ links to agent $j$ depends not only on the popularity of agent $j$, but also on the gregariousness of agent $i$, and the types of agents $i$ and $j$. Next, in the following Theorem we study the impact of the exogenous network parameters on the agents' popularity growth rates. 
\begin{thm}
{\it (Popularity growth in non-homophilic societies)} For a network with $h_{k} = 0, \forall k \in \Theta$, the popularity growth rates are given as follows:
\begin{itemize}
\item For $\gamma = 0$, the popularity of an agent $i$ grows {\it sublinearly} with time, i.e. $\mathbb{E}\left\{\mbox{deg}_{i}^{-}(t)\right\}$ is $O\left(t^{\frac{\bar{L}-1}{\bar{L}}}\right)$, where $\bar{L} = \sum_{k \in \Theta} p_{k} L^{*}_{k}(k,0)$, and the EPAT grows super-linearly with the degree of popularity, i.e. $T^{p}_{i}(d)$ is $O\left(d^{\frac{\bar{L}}{\bar{L}-1}}\right)$.        
\item For $\gamma = 1$, the popularity of an agent $i$ grows {\it logarithmically} with time, i.e. $\mathbb{E}\left\{\mbox{deg}_{i}^{-}(t)\right\}$ is $O\left(\bar{L}\,\log(t)\right)$, and the EPAT grows exponentially with the degree of popularity, i.e. $T^{p}_{i}(d)$ is $O\left(e^{\frac{d}{\bar{L}}}\right)$. \IEEEQEDhere          
\end{itemize} 
\end{thm}

This Theorem demonstrates the impact of opportunism and gregariousness on popularity accumulation. The popularity of opportunistic agents grows sublinearly with time, and the sublinearity exponent depends on the ``average" gregariousness of all types of agents in the society. On the other hand, if agents are not opportunistic, their popularity will grow only logarithmically with time. Thus, opportunism plays a fundamental role in determining the popularity growth rate. It is due to the emerging preferential attachment (which is promoted by agents' opportunism) that the popularity follows a sublinear growth over time. However, since the society is non-homophilic, the meeting process is not doubly preferential and the growth rates of all agents are the same. In fact, it is only the agent's age in the network that decide its expected popularity level as we show in the next Corollary.
\begin{crlry}
{\it (Superiority of older agents in non-homophilic societies)} For a network with $h_{k} = 0, \forall k \in \Theta$, $T^{p}_{c}(i,j,\theta_{i}, \theta_{j}) = \infty, \forall j > i$. That is, the expected popularity $\mathbb{E}\left\{\mbox{deg}_{i}^{-}(t)\right\}$ of an agent $i$ at any time step dominates the expected popularity of all younger agents.  \IEEEQEDhere
\end{crlry}

This Corollary says that popularity crossover does not occur in non-homophilic societies. This is a direct consequence of agents being indifferent to other agents' types, thus it is only the birth dates that distinguishes the agents. In the following Theorem, we compute the popularity crossover time for the same agent for $\gamma = 0$ and $\gamma = 1$, and show that opportunism leads to popularity gains in the long-run. 
\begin{thm}
{\it (Opportunism is good in the long-run)} For a network with $h_{k} = 0, \forall k \in \Theta$, the popularity crossover time $T^{p}_{c}(i,i,\theta_{i}, \theta_{i},\gamma,\gamma^{'})$ is always finite and can be approximated by  
\[T^{p}_{c}(i,i,\theta_{i}, \theta_{i}, 1, 0) \approx i \times \left(-\bar{L} \,\, \mathcal{W}_{-1}\left(\frac{1}{\bar{L}}e^{\frac{-1}{\bar{L}}}\right)\right)^{\frac{\bar{L}}{\bar{L}-1}},\]
where $\mathcal{W}_{-1}(.)$ is the lower branch of the Lambert W function \cite{ref372}. \IEEEQEDhere
\end{thm}

The popularity crossover time increases linearly with the agent's birth date, and grows as $O\left(\bar{L} \, \log(\bar{L})\right)$ with the agents' average gregariousness. Thus, younger and more gregarious agents need to wait longer to harvest the popularity gains attained by opportunism. Fig. 3 shows the expected popularity growth over time for agents born at $t$ = 10, 20, and 30. Solid lines are the logarithmically growing popularity if agents are not opportunistic, while dashed lines correspond to the sublinearly growing expected popularity for the opportunistic agents, and it can be seen that a finite crossover time exists for all such agents.

In the results above, we focused on non-homophilic societies, where meetings are not doubly preferential and different types go through the same popularity evolution process. In the following Theorem, we evaluate the popularity growth rates in homophilic societies.  
\begin{thm}
{\it (Popularity growth in homophilic societies)} For a network with $h_{k} = 1, \forall k \in \Theta$, the popularity growth rates are given as follows:
\begin{itemize} 
\item For $\gamma = 0$, the popularity of an agent $i$ grows {\it sublinearly} with time, i.e. $\mathbb{E}\left\{\mbox{deg}_{i}^{-}(t)\right\}$ is $O\left(t^{b}\right)$, and the EPAT grows super-linearly with the degree of popularity, i.e. $T^{p}_{i}(d)$ is $O\left(d^{\frac{1}{b}}\right)$, where $b = \frac{L^{*}_{\theta_{i}}(\theta_{i},0)-1}{L^{*}_{\theta_{i}}(\theta_{i},0)}$.        
\item For $\gamma = 1$, the popularity of an agent $i$ grows {\it logarithmically} with time, i.e. $\mathbb{E}\left\{\mbox{deg}_{i}^{-}(t)\right\}$ is $O\left(L^{*}_{\theta_{i}}(\theta_{i},0) \, \log(t)\right)$, and the EPAT grows exponentially with the degree of popularity, i.e. $T^{p}_{i}(d)$ is $O\left(e^{b d}\right)$, where $b = \frac{1}{L^{*}_{\theta_{i}}(\theta_{i},0)}$.   \IEEEQEDhere       
\end{itemize} 
\end{thm}
       
It is clear that for homophilic societies, the popularity growth rate are the same as those computed in Theorem 4, yet the sublinearity exponent for opportunistic agents' popularity growth, and the scaling factor for the logarithmic popularity growth of non-opportunistic agents depend on the gregariousness of each type of agents. Thus, agents that are more gregarious experience faster growth rates, which implies that doubly preferential meeting process promotes the popularity of {\it older} and more {\it gregarious} agents by the effects of preferential attachment and homophily respectively as we show in the following Corollary.     	
\begin{crlry}
{\it (Younger and more gregarious agents are more popular than older and less gregarious agents)} For a network with $h_{k} = 1, \forall k \in \Theta$, $T^{p}_{c}(i,j,\theta_{i}, \theta_{j}) < \infty, \forall j > i$ if and only if $L^{*}_{\theta_{j}}(\theta_{j},0)>L^{*}_{\theta_{i}}(\theta_{i},0)$.  \IEEEQEDhere
\end{crlry}

Thus, the fraction of type-$k$ agents in the society does not affect its chances of getting popular. In fact, it is the gregariousness of type-$k$ agents, in addition to the structural opportunism, that control the popularity evolution. Structural opportunism leads to preferential attachment, which gives a cumulative advantage to older agents to get more popular, while homophily creates a doubly preferential meeting process, which together with the heterogeneity in the agent's gregariousness can promote the popularity of certain types of agents. The net effect can be that, unlike the case of non-homophilic societies, a younger agent can on average get more popular than an older agent if the younger agent belongs to a more gregarious type. Fig. 4  shows that the popularity of a younger, but more gregarious agent born at $t=30$ exceeds that of an older agent born at $t=20$. The results of Theorem 6 can be used to interpret the interesting empirical results on citation networks in \cite{ref373}, where it has been shown that there is a correlation between the number of references per paper and the total number of citations in a certain scientific field. We quote the following conclusion from the report in \cite{ref577}, which is based on a statistical analysis of Thomson Reuters' Essential Science Indicators database: ``{\it One might think that the number of papers published or the population of researchers in a field are the predominant factors that influence the average rate of citation, but it is mostly the average number of references presented in papers of the field that determines the average citation rate.}" Such conclusion is in perfect agreement with Theorem 6 (and Corollary 3), where we predict that for the inherently homophilic citation networks, the popularity of researchers in different fields (total citation rate) is governed by their ``gregariousness" (number of references per paper), and not by the type distribution (number of papers/researchers). We know from \cite{ref577} that mathematics papers typically list few references, whereas those in molecular biology display extensive citations. Thus, molecular biologists are more ``gregarious" than mathematicians, and one would expect that younger molecular biologists can, on average, get more popular than mathematicians. To illustrate this, we have collected the publicly available Google Scholar citation data for all molecular biologists who started publishing papers in the year 2000, and the citation data for mathematicians who started publishing in 1993. As predicted by Theorem 6 and Corollary 3, we show in Fig. 5\footnote{Note that popularity growth in Fig. 5 is super-linear due to the non-stationary arrival process and gregariousness, which we will incorporate in our future extensions for the model.} that a popularity crossover occurs for the ``young" molecular biologists in the year 2006.               

\begin{figure}[t!]
    \centering
    \includegraphics[width=3 in]{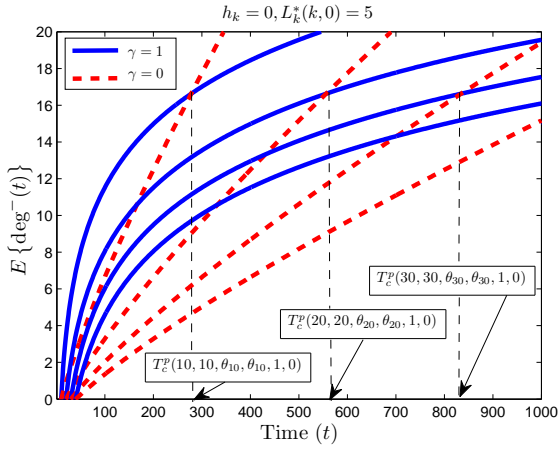}     
		\captionsetup{font= small}
    \caption{Popularity crossover times for agents with and without structural opportunism. Agents are born at $t$ = 10, 20, and 30.}
\end{figure}
\begin{figure}[t!]
    \centering
    \includegraphics[width=3 in]{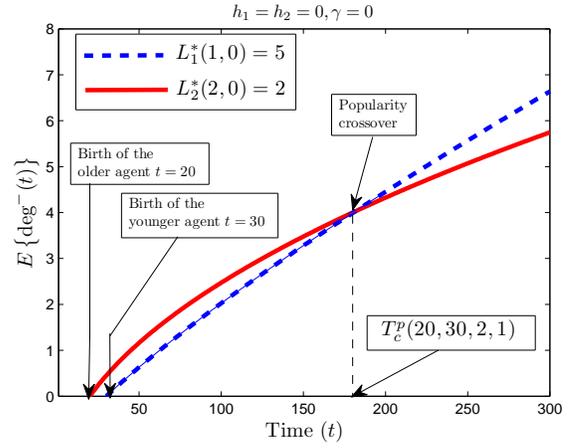}     
		\captionsetup{font= small}
    \caption{Popularity crossover times in homophilic societies.}
\end{figure}
\begin{figure}[t!]
    \centering
    \includegraphics[width=3 in]{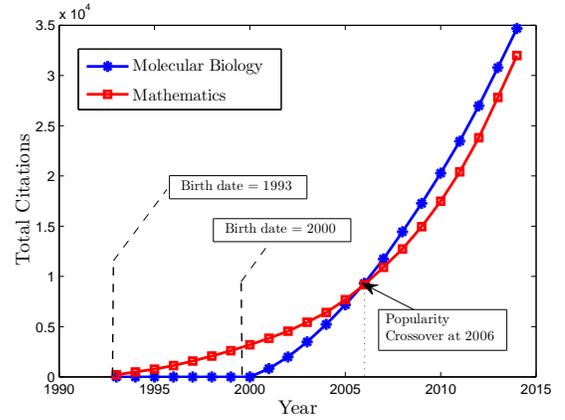}     
		\captionsetup{font= small}
    \caption{Popularity crossover for molecular biology researchers with respect to mathematicians.}
\end{figure}

\section{Discussion and Future Extensions}
A comprehensive theoretical model for the evolution of social networks was presented, and the associated analysis for agent-level link formation and acquisition was carried out. In this section, we discuss various uses and applications of our model. In particular, the model can be used, in addition to providing insights into the network evolution process, to carry out agent and network-level predictions, and designing policies that guide network formation.   

\subsection{Network Prediction}

Our model can make useful predictions about the network structure and individuals popularity, time to acquire friendships etc. and how these change when the interacting agents exhibit different characteristics. Moreover, our model predicts the emergence of preferential attachment, which is widely observed in many real-world networks \cite{ref4}, as a result of the agents' dynamic meeting process and linking actions. In contrast, data driven approaches \cite{ref10}-\cite{ref24} \cite{ref31}-\cite{ref36} cannot provide such network characterizations and predictions unless they have access to an abundance of relevant data. Moreover, the proposed microfoundational model enables us to determine what networks will form and how will they evolve when the agents characteristics are different and to understand what would be different if the agents would have different characteristics. Such detailed comparisons, analysis and counterfactuals are not possible based on data-driven approaches because this would require access to enormous amounts of data and, even more importantly, access to networks that cannot be monitored and may not even exist (yet). For instance, Leskovec {\it et. al} \cite{ref39} characterized the friendship acquisition time, where a parameterized model is constructed with the aid of a large data set of temporal {\it node} (agent) arrivals and {\it edge} (link) creation times for Linkedin, Yahoo! Answers, and Flickr. The {\it time gap} between the creation of two links by the same node is shown to fit an exponential distribution, and estimates for the exponential distribution parameter for different networks are provided. While \cite{ref39} estimates the exponential distribution parameters for different networks, it implicitly assumes that all types of agents go through the same experience, and in addition cannot explain why different networks have different time gap statistics, and what will happen if different parameters are changed. In contrast, by calibrating the values of the exogenous parameters, our model can explain why the LFT of agents in different networks would differ, how is it affected by the exogenous parameters, and can in addition handle counterfactuals.  

\subsection{Policy design}
Our model can be used to not only carry out link predictions, but it can also be used to analyze and design new policies that can modify and guide the formation and evolution of networks. An example for a policy is to influence the agents' meeting process at every time step given the observed step graph at the previous time step. This corresponds to the {\it link recommendation} problem \cite{ref575}, i.e. suggesting friends for the agent over time. Such process of guiding link creation are of interest to network planners and designers since in many OSN, such as Facebook, the friends recommendation system is responsible for a significant fraction of link creations. The policy design problem can have different objectives such as: controlling the level of popularity of different types of agents, minimizing the LFT, or controlling the community structure.

\end{document}